\documentclass[lettersize,journal]{IEEEtran}
\usepackage{colortbl}
\definecolor{mygray}{gray}{.9}
\usepackage{cite}
\usepackage{color}
\usepackage{amsfonts,amssymb}
\usepackage{bm}
\usepackage{amsmath}
\usepackage{multirow}
\usepackage{algorithmic}
\usepackage{graphicx}
\usepackage{wrapfig}
\usepackage{subfigure}
\usepackage{float}
\usepackage{textcomp}
\usepackage{graphicx} 
\usepackage{epstopdf}
\usepackage{array}
\usepackage[thmmarks,amsmath]{ntheorem}
\newtheorem{mythm}{Theorem}
\newtheorem{mydef}{Definition}
\newtheorem{proposition}{Proposition}
\newtheorem{assumption}{Assumption}
\newtheorem{lemma}{Lemma}
\newtheorem{remark}{Remark}

\usepackage[title]{appendix}

\usepackage{array}
\usepackage{enumerate}
\usepackage{booktabs}
\usepackage{algorithm}
\usepackage{algorithmic}
\usepackage{setspace}
\usepackage{float}
\usepackage{cases}

\usepackage{mathrsfs}

\begin{document}

\title{Deviation Between Team-Optimal Solution and Nash Equilibrium in Flow Assignment Problems
}

\author{ Gehui Xu$^{1}$, Ting Bai$^{2}$, Andreas A. Malikopoulos$^{2}$, and Thomas Parisini$^{3}$
\thanks{*This work was supported in part by NSF under Grants CNS-2401007, CMMI-2348381, IIS-2415478, and in part by MathWorks.}
\thanks{$^{1}$G. Xu is with the Department of Electrical and Electronic Engineering, Imperial College London, SW7 2AZ London, U.K. E-mail: {\tt\small g.xu@imperial.ac.uk}}
\thanks{$^{2}$T. Bai and A. A. Malikopoulos are with the Information and Decision Science Lab, School of Civil $\&$ Environmental Engineering, Cornell University, Ithaca, New York, U.S.A. E-mails: \{{\tt\small tingbai, amaliko\}@cornell.edu}}
  \thanks{$^{3}$Thomas Parisini is with  the Dept. of Electrical and Electronic Engineering,
	Imperial College London, London SW7 2AZ, UK, and also with the Dept. of Electronic Systems, Aalborg University, Denmark and with the  Dept.
	of Engineering and Architecture, University of Trieste, Italy.
        E-mail: {\tt\small t.parisini@imperial.ac.uk}}
}

\maketitle
{
\begin{abstract} 
We investigate the relationship between the team-optimal solution and the Nash equilibrium (NE) to assess the impact of strategy deviation on team performance. As a working use case, we focus on a class of flow assignment problems in which each source node acts as a cooperating decision maker (DM) within a team that minimizes the team cost based on the team-optimal strategy.
In practice, some selfish DMs may prioritize their own marginal cost and deviate from NE strategies, thus potentially degrading the overall performance.
To quantify this deviation, we explore the deviation bound between the team-optimal solution and the NE in two specific scenarios: (i) when the team-optimal solution is unique and (ii) when multiple solutions do exist. This helps DMs analyze the factors influencing the deviation and adopting the NE strategy within a tolerable range.
Furthermore, in the special case of a potential game model, we establish the consistency between the team-optimal solution and the NE. Once the consistency condition is satisfied, the strategy deviation does not alter the total cost, and DMs do not face a strategic trade-off.
Finally, we validate our theoretical analysis through extensive simulation trials.
\end{abstract}}


\section{Introduction}

Routing in networks for data transmission and logistics management has been widely studied across various domains, including wireless communication \cite{orda1993competitive}, road traffic \cite{bai2025routing}, and freight transportation \cite{crainic2000service}. In particular, increasing attention has been given to a class of
 flow assignment problems \cite{aicardi1989decentralized,aicardi1990decentralized, leung1999flow}. These problems aim to determine the optimal allocation of traffic across multiple available routes from each source node to the destination node  to achieve a desired objective, such as minimizing the total transmission cost of the network.

Team theory provides a well-established framework for modeling flow assignment problems from a cooperative perspective \cite{radner1962team, aicardi1989decentralized, aicardi1990decentralized}. In this context, source nodes act as decision makers (DMs) within a team organization. Based on available information, such as measured input flow values and network topology, each DM allocates its received flows across multiple paths leading to the destination. While each DM independently decides its routing strategy, they share the common objective of minimizing the total cost (i.e., the team cost)
which is influenced by their joint decisions \cite{radner1962team,  aicardi1989decentralized, aicardi1990decentralized, Dave2021a}. The information structure in these problems can be categorized as  {\it static} or {\it dynamic}~\cite{Malikopoulos2021,Dave2021nestedaccess}. In static team problems, the information available to each DM is 
not affected by the decisions of others~\cite{aicardi1989decentralized, aicardi1990decentralized}. In contrast,  dynamic team problems mean that the information of at least one DM is affected by the decisions of others~\cite{Malikopoulos2021}. In both static and dynamic team problems, the team outcome is characterized by the team-optimal solution~\cite{radner1962team,zoppoli2020neural,Malikopoulos2021}, defined as a strategy profile under which the overall team performance cannot be improved by changing the strategy of one or more members~\cite{Dave2021a}. 


However and inevitably, there may be  selfish DMs in a team  who prioritize their own costs thus being reluctant
to adopt the team-optimal strategy~\cite{altman2007evolutionary,haurie1985relationship}. This is because achieving the team optimum often requires DMs to sacrifice their own benefits, such as tolerating longer transport time in logistics to reduce overall network costs~\cite{shivshankar2014evolutionary} or consuming additional energy in wireless networks to facilitate data transmission~\cite{ orda1993competitive}. 
Additionally, due to limited computing and communication resources, DMs may find it difficult to access complete network information needed to minimize the team's total cost and instead make decisions based on local information. As a result, the flow assignment problem could  also be formulated as a non-cooperative game~\cite{altman2007evolutionary,haurie1985relationship}. In this context, DMs focus on minimizing their individual costs based on the Nash equilibrium (NE)~\cite{bacsar1998dynamic}, which represents a strategy profile where no DM can improve its profit by changing its strategy, given the strategies of others.

As selfish DMs prioritize minimizing their individual costs rather than the team's overall cost,
 deviating from the team-optimal strategy to the NE strategy may compromise the overall team performance.
However, it is worth mentioning that if the team-optimal solution and the NE are of little difference, then such a strategy
deviation will not significantly impact the collective interests of all DMs, as the total cost increase remains acceptable. Moreover, in certain special cases where the NE coincides with the team-optimal solution, regardless of whether the DMs deviate in their strategies, the total costs remain unchanged.
 
Therefore, the main objective of this paper is to explore the relationship between the team-optimal solutions and the NE to evaluate how strategy deviations impact the team's overall cost. 
We focus on a simple flow assignment problem and formulate it as a static team problem and a non-cooperative game problem, respectively.
We first provide an upper bound on the deviation between the team-optimal solution and the NE in two cases: (i) when the team-optimal solution is unique and (ii) when multiple solutions exist.
Next, we establish the consistency between the team-optimal solution and the NE in a specific non-cooperative game setting, namely, a potential game.
Finally, we present simulation studies on two logistics transportation examples to validate our theoretical analysis.

\section{Flow assignment problem}

In this section, we begin with a routing example in a logistics system. Next, we present the mathematical formulation of the flow assignment problem, modeling it both as a static cooperative team problem and a non-cooperative game.
\vspace{-0.3cm}
\subsection{A freight logistics problem}\label{sec:example}

As a motivating example, consider a freight logistics problem where multiple independent shippers 
share a transportation network \cite{crainic2000service}.    The shippers need to transport goods to a common destination (e.g., a warehouse or a distribution center) through multiple available transportation paths, which are connected by multiple edges. 
The transportation cost of each edge depends on the total volume of goods transported by all shippers using that edge. 
{ We assume all shippers belong to the same carrier and collectively minimize the total transportation cost of the carrier.}
\vspace{-0.3cm}
 \subsection{Cooperative team decision-making problem}
Based on the example in Section \ref{sec:example}, we now introduce the mathematical formulation of the flow assignment problem in a general form. 
Team theory \cite{radner1962team,Malikopoulos2021,aicardi1989decentralized,aicardi1990decentralized} is a mathematical formalism that can be used to model this problem. It was developed to provide a rigorous mathematical framework for cooperating members in which all members have the same objective, yet different information.   In this context, we consider a ``team" consisting of a number of members that cooperate to achieve a common objective. 
The underlying structure to model a team decision problem consists of  (1) a number of members of the team; (2) the decisions of each member; (3) the information available to each member, which may be different; (4) a global objective, which is the same for all members; and (5) the existence, or not, of communication between team members. 

\begin{figure}[ht]
	\centering	
    \vspace{-0.2cm}
	\includegraphics[scale=0.11]{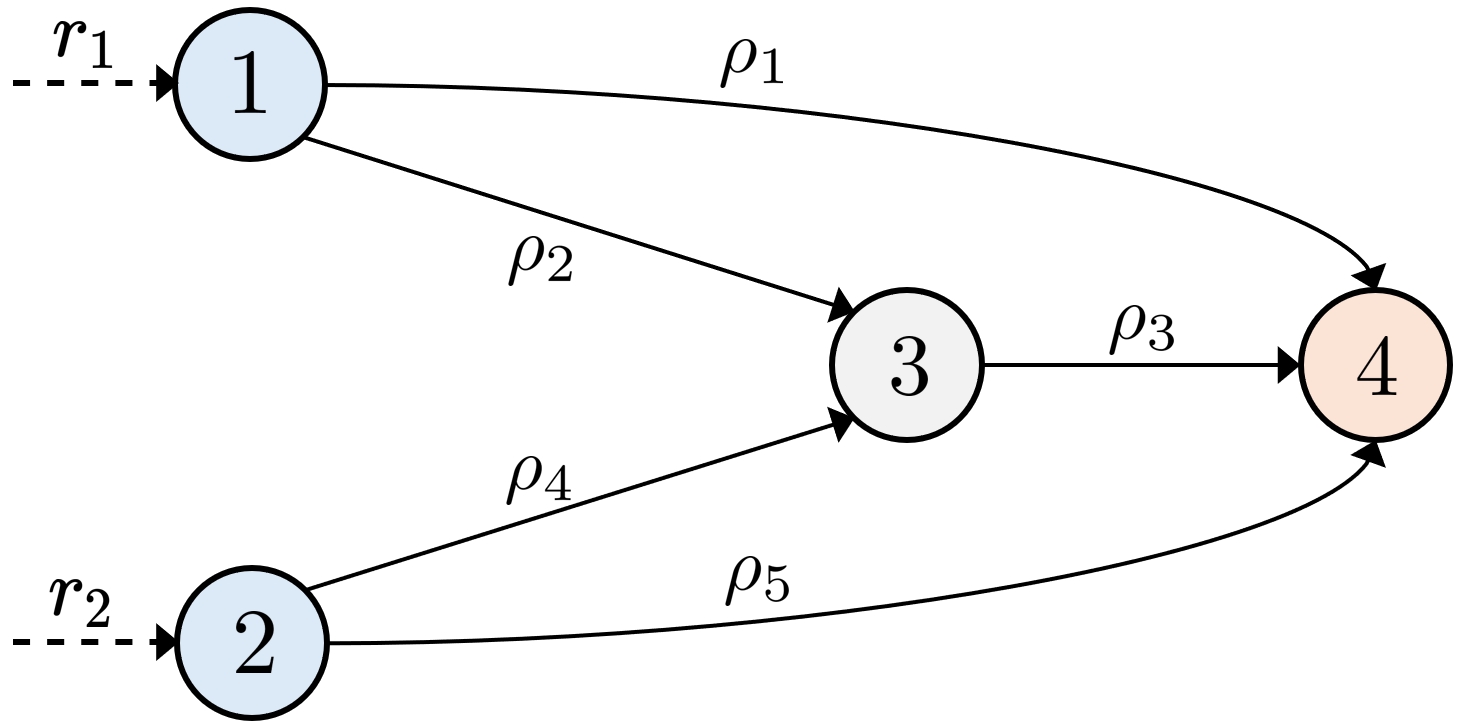}\\
 \vspace{-0.2cm}
	\caption{Flow assignment problem with two source nodes.
    Source nodes 1 and  2 in blue represent  DM$_1$ and DM$_2$. 
    Node 3 in gray serves as a transmitting node, and node 4 in red is the destination. The parameters $\rho_1,\dots,\rho_5$ represent the weighting coefficients.
    }
	\label{fig127}
	\vspace{-0.4cm}
\end{figure}

In this problem, the routing nodes have decision-making capabilities and are considered cooperating members of a team organization.
Specifically,
 we model the network as a directed graph
       $\mathcal{G}(\mathcal{V},\mathcal{E})$,  where $\mathcal{V}=\{1,\dots,M\} ,M\in\mathbb{N}$, is the node set  and $\mathcal{E}\subseteq \mathcal{V} \times \mathcal{V}$ is the  edge set. 
Moreover, there are $N$ ($N<M$, $N\in\mathbb{N}$) source nodes and a single destination node. Each source node corresponds to a shipper. Let $\mathcal{N}=\{1,\dots,N\}$ denote the  set of  source nodes.
The routing decisions are made only at the source nodes by the DMs. For illustration, in Fig.~\ref{fig127} we show a flow assignment problem with two DMs.

{For $i\in \mathcal{N}$, let $r_i\in\mathbb{R}$ be 
the flow entering  
DM$_i$,
which represents the amount of goods received by shipper~$i$ from an external supplier.
Each DM
$i$ has $ P_i\in\mathbb{N}$  available paths to the destination, connected by multiple edges.
Let $L=|\mathcal{E}|\in\mathbb{N}$ be the number of edges in the network. We introduce an indicator function $\sigma(l, k, i)$,
which takes the value of 1 if edge  $l$ (for $l=1,\dots,L$) belongs to  $k$-th  path (for $k=1,\dots,P_i$)  from DM$_i$ to the destination, and 0 otherwise.  { Each DM 
$i$ makes a routing decision 
$u_i=\operatorname{col}\{u_i^{k}\}_{k=1}^{P_i}\in \mathbb{R}^{P_i}$ 
to allocate the external flow 
$r_i$ among the $P_i$ available paths.  Here,  
$u_i^{k}\in\mathbb{R}$ denotes the portion of external flow  $r_i$ assigned to path $k$,
representing the amount of goods allocated to transport route $k$. Note that the value of $u_i^{k}$ is the same for all links that belong to path $k$. 
 The local strategy set of DM$_i$ is defined by 
\begin{align*}
 \Xi_i= \left \{u_i\in \mathbb{R}^{P_i}: \sum_{k=1}^{P_i} u_i^{k}=r_i, 
 u_i^k \geq 0, \text{for}\,k=1,\dots,P_i
 \right \}. \end{align*}
 }
 
 Let $   \boldsymbol{\Xi}=\prod_{i=1}^{N}\Xi_{i} \subseteq \mathbb{R}^{\sum_{i=1}^N P_i} $ be the joint strategy set for all DMs, } 
$\boldsymbol{u}=\operatorname{col}\{u_i\}_{i=1}^N\in \boldsymbol{\Xi}$  be the strategy profile for all DMs, and $\boldsymbol{u}_{-i}=\operatorname{col}\{u_1,\dots,u_{i-1},u_{i+1},\dots,u_{N}\}$ be the strategy profile for all DMs except DM$_i $.

All DMs share a common objective of minimizing the total cost of the entire network.
The cost is determined based on edge costs, which in turn depend on the total flow allocated by all DMs. 
Let 
\vspace{-0.2cm}
\begin{align*}
z_l(\boldsymbol{u})=\sum_{i=1}^N \sum_{k=1}^{P_i} u_i^k \sigma(l, k, i)
  \end{align*}
be the total flow function on edge $l$, and let 
$f_{l}(z_l)$ be the transmission cost function on edge $l$. One common form of $f_{l}$ is quadratic, i.e.,
\vspace{-0.2cm}
\begin{align*}
f_{l}(z_l(\boldsymbol{u}))=z_l^2(\boldsymbol{u})={\left [ \sum_{i=1}^N \sum_{k=1}^{P_i} u_i^k \sigma(l, k, i) \right ]}^2.
\end{align*}
Consider $\rho_{l}>0$ as the weighting coefficient related to edge $l$, 
representing the transportation cost per unit of goods on that edge.
Then the total cost function of the entire network is 
\begin{equation}\label{totalcost}
\mathcal{C} (\boldsymbol{u})= 
\sum_{l=1}^L \rho_{l} \cdot f_{l} (z_l(\boldsymbol{u})).\vspace{-0.2cm}
\end{equation}
In this context, 
each DM independently determines its routing strategy $u_i$ and collaborates in minimizing $\mathcal{C}$. The team decision-making problem is formulated as follows:
\begin{equation}\label{team_function}
\begin{aligned}
&\min_{u_1,\dots,u_N}\; \mathcal{C} (u_1,\dots,u_N) \\
& \mathrm{s.\,t.}
\;  
u_i\in \Xi_i,\;
i\in\mathcal{N}.
\end{aligned}
\end{equation}

{Since $r_i$ for $i\in \mathcal{N}$ are independent parameters, each DM  receives this information and makes decisions independently. This indicates that the information received by each DM is not influenced by the decisions of others \cite{aicardi1989decentralized,aicardi1990decentralized,zoppoli2020neural}. 
 Thus, the team problem under consideration follows a static information structure with independent available information~\cite{zoppoli2020neural}.   Referring to \cite{aicardi1989decentralized,aicardi1990decentralized}, we have the following assumption for the decision-making of DMs}.
\begin{assumption}\label{assum1}
For  $i\in\mathcal{N}$, the routing decision $u_i$ is determined based on the complete knowledge of the network topology $\mathcal{G}$ and the input flow $r_i$. 
\end{assumption} 
The corresponding outcome is described by a team-optimal solution \cite{aicardi1989decentralized,aicardi1990decentralized,zoppoli2020neural}, formally defined next.

\begin{mydef}[Team-optimal solution]
A strategy profile $\boldsymbol{u}^*=\operatorname{col}\{u_1^*, u_2^*, \dots, u_N^*\}\in \boldsymbol{\Xi}$ is a team-optimal solution if 
\begin{align*}
    \mathcal{C}(\boldsymbol{u}^*)\leq  \mathcal{C}(\boldsymbol{u}), \; \forall \boldsymbol{u} \in \boldsymbol{\Xi}.
\end{align*}
\end{mydef}
\subsection{Non-cooperative game}\label{Non-cooperative game}

 Next, we reformulate the flow assignment problem as a non-cooperative game.

Let $\{\mathcal{N}, \{\Xi_i\}_{i\in\mathcal{N}}, \{\mathcal{C}_i\}_{i\in\mathcal{N}}\}$ be a non-cooperative game, where $\mathcal{C}_i(u_{i}, \boldsymbol{u}_{-i}):\mathbb{R}^{\sum_{i=1}^N P_i}\rightarrow \mathbb{R} $ denotes DM$_i$'s payoff function, representing its selfish interest; it is associated with the transmission cost of its allocated flow across the network. 
Let $g_{l}(z_l(\boldsymbol{u}))$ be the unit flow cost function, which is determined by all DMs using edge $l$.
One common
form of $g_{l}$ is the linear form \cite{altman2007evolutionary},  i.e.,
\vspace{-0.2cm}
\begin{align*}
g_{l}(z_l(\boldsymbol{u}))=a_l+b_l z_l(\boldsymbol{u})=a_l+b_l\sum_{i=1}^N\sum_{k=1}^{P_i} u_i^k \sigma(l, k, i),
\end{align*}
where $a_l\geq 0$ is the fixed cost and $b_l>0$ is the congestion cost. 
In a non-cooperative setting, each DM only cares about
its own proportional share on edge $l$, given by  
$(\sum_{k=1}^{P_i} u_i^k \sigma(l, k, i))\cdot g_{l}(z_l(\boldsymbol{u})) $.
 Then the individual payoff $\mathcal{C}_i(u_{i}, \boldsymbol{u}_{-i})$ for each DM$_i$ is 
\begin{equation}\label{payoff1}
\begin{aligned}
{\mathcal{C}}_i(u_{i}, \boldsymbol{u}_{-i})\!=\!  \sum_{l=1}^L \rho_{l} \omega(l,  i) \left [ \sum_{k=1}^{P_i} u_i^k \sigma(l, k, i)\right ] g_{l}(z_l(\boldsymbol{u})),
\end{aligned}
\end{equation}
where $\omega(l,  i)$ is an indicator function which simply gives~1 if edge  $l$ related with DM$_i$, and is 0 otherwise.
{In practice, DMs often prioritize minimizing their own costs due to self-interest rather than cooperatively optimizing the overall network performance. 
 Given $\boldsymbol{u}_{-i}$, DM$_i$ aims to solve}
\begin{equation}\label{f1}
	\min \nolimits_{u_{i} \in \Xi_{i}} \mathcal{C}_i\left(u_{i}, \boldsymbol{u}_{-i}\right). \quad  
\end{equation}

The outcome of game \eqref{f1} is characterized by the NE \cite{nash1951non}.
\begin{mydef}[NE]\label{d2}
	A profile $ \boldsymbol{u}^{\Diamond}=\operatorname{col}\{u_{1}^{\Diamond}, \dots, u_{N}^{\Diamond} \} \in \boldsymbol{\Xi}
    $ is said to be an  NE of game $\{\mathcal{N}, \{\Xi_i\}_{i\in\mathcal{N}}, \{\mathcal{C}_i\}_{i\in\mathcal{N}}\}$ if 
	\begin{equation*}\label{ne}
		\mathcal{C}_i\left(u_{i}^{\Diamond}, \boldsymbol{u}_{-i}^{\Diamond}\right) \leq \mathcal{C}_i\left(u_{i}, \boldsymbol{u}_{-i}^{\Diamond}\right),\;\forall i \in \mathcal{N}, \forall u_{i}\in \Xi_{i}. 
	\end{equation*}
\end{mydef}
NE characterizes an outcome where no DM has an incentive to unilaterally deviate from their chosen strategy, provided that the strategies of all others remain unchanged. 

{\begin{remark}
The team decision-making problem can be analyzed from a game-theoretic perspective, referred to as a team game, where all DMs share the same preferences and seek to minimize a common cost function $\mathcal{C}$ \cite{bacsar1998dynamic}. In this context, the NE corresponds to person-by-person optimality within the team, a sub-optimal solution concept than the team-optimal solution, and satisfies
\begin{align*}
    \mathcal{C}(u_i^{\Diamond},\boldsymbol{u}_{-i}^\Diamond)&\leq  \mathcal{C}(u_i,\boldsymbol{u}_{-i}^\Diamond), \; \forall i\in \mathcal{N}, u_i\in \Xi_i.
\end{align*}
Clearly, any team-optimal solution is also an NE, although the converse does not necessarily hold.
 
\end{remark}}

Define  $G(\boldsymbol{u})=\nabla_{\boldsymbol{u}}\mathcal{C}(\boldsymbol{u})=\operatorname{col}\{\nabla_{{u_1}}\mathcal{C}(\cdot,u_{-1}),\dots,$ $\nabla_{{u_N}}\mathcal{C}(\cdot,u_{-N})\}
$ as the gradient map of \eqref{totalcost}  and  $F(\boldsymbol{u})=\operatorname{col}\{\nabla_{{u_1}}\mathcal{C}_1(\cdot,u_{-1}),$ $\dots,\nabla_{{u_N}}\mathcal{C}_N(\cdot,u_{-N})\}
$ as the
pseudo-gradient map of \eqref{payoff1}.
We impose the following assumption.

\begin{assumption}\label{assum2}
	$\ $
	\begin{enumerate}[(1)]
    \item 
    The function $\mathcal{C}(\boldsymbol{u})$ is convex
and continuously differentiable in $\boldsymbol{u}$. 
For  $i\in\mathcal{N}$, the function $\mathcal{C}_i(u_i,\boldsymbol{u}_{-i})$ is convex
and continuously differentiable in  $u_i$. 
\item The operators $G(\boldsymbol{u})$ and $F(\boldsymbol{u})$  are monotone. 
\end{enumerate}
\end{assumption}

 Assumption \ref{assum2} can guarantee the existence of a team-optimal solution and an NE by using variational inequality \cite{facchinei2003finite}. Also,  Assumption \ref{assum2} does not restrict the uniqueness of the team-optimal solution and NE.
 The following results verify the existence of a team-optimal solution and an NE \cite{facchinei2003finite}. 
	\begin{lemma}\label{l1}
			Given Assumption \ref{assum2},  there exists a team-optimal solution  $\boldsymbol{u}^{*}$.
		\end{lemma}
\begin{lemma}\label{l2}
			Given Assumption \ref{assum2},  there exists an NE  $\boldsymbol{u}^{\Diamond}$.
		\end{lemma}

{ In prior works \cite{aicardi1989decentralized,aicardi1990decentralized},  DMs adopt team-optimal strategies based on Assumption 1.} 
However, this assumption may not always hold in practice. DMs may prioritize their own interests, being reluctant to compromise their performance or share information. Consequently, some DMs may deviate from the team-optimal solution, adopting the NE strategy to minimize their own costs, which could potentially reduce overall team performance. 
Such a strategy deviation may compromise the collective performance of the team. 
 For example, in the freight transportation case, some shippers select the cheapest route for themselves, resulting in congestion and increased costs for others, which causes the total system cost to exceed that of the team-optimal solution.

Nevertheless, 
if the team-optimal solution
and the NE are nearly identical, such a strategy deviation has minimal impact. In special cases where they coincide, total costs remain unchanged regardless of individual strategy shifts.
To this end, we seek to address the following questions in this paper:
\begin{itemize}
    \item Does there exist a deviation bound between the team-optimal solution and the NE? 
    \item In which scenario is the team-optimal solution and the NE consistent?
\end{itemize}
{\begin{remark}
A concept that may be confused with the team‐optimal solution is the social optimum, which is well‐known in the game literature \cite{koutsoupias1999worst}.
 Both seek to maximize total welfare, but in general, these two optima do not coincide. 
 The team solution adopts a unified perspective to optimize  a given team objective $\mathcal{C}(\boldsymbol{u})$ and selects the joint decision that yields the lowest team cost.
A social optimum, by contrast, respect each individual’s preference and aims to minimize $\sum_{i=1}^N\mathcal{C}_i(\boldsymbol{u})$. The inefficiency of NE strategies relative to the social optimum has been extensively studied through the concept of the price of anarchy \cite{koutsoupias1999worst}, whereas the inefficiency relative to the team optimum has not been thoroughly explored.
\end{remark}}


\section{Deviation bound between team-optimal solution and NE}

In this section, we explore the deviation bound between a unique team-optimal solution and a unique NE.   {Based on the convexity and continuity condition in Assumption \ref{assum2}, the team-optimal solution and the NE are equivalent to the first-order stationary points of their respective problems. This allows us to study their deviation based on gradient information.}
 Let us denote the gradient difference 
 \begin{align}
      \theta(\boldsymbol{u})=G(\boldsymbol{u})-F(\boldsymbol{u}) \nonumber
 \end{align}
as a perturbation term.
Additionally, we define the compact set  $\boldsymbol{\Theta}=\{\boldsymbol{u}\in\boldsymbol{\Xi}: \|\boldsymbol{u}-\boldsymbol{u}^*\|< q\}$, where $q$ is a positive constant. 
The following result provides a deviation bound. 
The proof is presented in Appendix \ref{lea8}.

\begin{mythm}\label{t7}
		Given Assumption \ref{assum2}, suppose that  $G(\boldsymbol{u})$ is  $\kappa_1$-strongly monotone and $F(\boldsymbol{u})$ is  strongly monotone.
        If
     the perturbation term $\theta(\boldsymbol{u})$ satisfies   $\|\theta(\boldsymbol{u})\|\leq \delta <\kappa_1\cdot s\cdot q $
     for all $\boldsymbol{u}\in\boldsymbol{\Theta}$ and
     some constant $0<s<1$, 
         \begin{align}
         \|\boldsymbol{u}^*-\boldsymbol{u}^{\Diamond}\|\leq \frac{ \delta}{\kappa_1 s}.
     \end{align}
\end{mythm}


By Theorem~\ref{t7}, 
since $\kappa_1$ and $s$ are constants,
the  deviation  $\|\boldsymbol{u}^*-\boldsymbol{u}^{\Diamond}\|$ is influenced by a term proportional to $\|\theta(\boldsymbol{u})\|$. Since $\boldsymbol{\Xi}$ is compact and $\theta(\boldsymbol{u})$  is continuous, there exists a constant  $\delta$  such that $\theta(\boldsymbol{u})\leq \delta$.  Given any such $\delta$, we can always select  $q$ large enough to ensure $\|\theta(\boldsymbol{u})\|\leq \delta <\kappa_1 s q $, ensuring that the deviation $\|\boldsymbol{u}^*-\boldsymbol{u}^{\Diamond}\|$ remains bounded. Therefore, the deviations from the team-optimal solution do not grow unbounded under small perturbations. 
{ If $\theta(\boldsymbol{u})$ is sufficiently small, then $\|\boldsymbol{u}^*-\boldsymbol{u}^{\Diamond}\|$  remains small. This implies that DMs can tolerably adopt the NE strategy, as the resulting cost difference remains within an acceptable range.}

Next, we  relax the assumptions on \( G(\boldsymbol{u}) \) and \( F(\boldsymbol{u}) \) in Theorem~\ref{t7} and consider the case where the solution is not unique.
To quantify the deviation between solution sets, we employ the Hausdorff metric \cite{bronstein2008approximation}, which is one of the most common metrics for measuring the distance between two sets. 
The Hausdorff metric considers the greatest distance that a point in one set needs to travel to reach the other \cite{bronstein2008approximation}. 
The Hausdorff metric of two sets $ A,B\subseteq  \mathbb{R}^{\sum_{i=1}^N P_i} $ is defined as 
\begin{align}
H(A,B) = \max\{\sup\limits_{a\in A}\inf\limits_{b\in B}\|a-b\|,\sup\limits_{b\in B}\inf\limits_{a\in A} \|a-b\|\}.\nonumber
\end{align}
Let $\Upsilon_{TO}$ be the set of  team-optimal strategy profile $\boldsymbol{u}^{*}$ and $\Upsilon_{NE}$ be the set of  NE strategy profile $\boldsymbol{u}^{\Diamond}$. Let
$\varLambda_{TO}$ be the image set of $F(\boldsymbol{u}^{*})$ and $\varLambda_{NE}$ be the image set of $F(\boldsymbol{u}^{\Diamond})$. We characterize the deviation between the NE and team-optimal sets via their differences in the gradient space. 
The proof is in Appendix  \ref{lea3}.

\begin{mythm}\label{c7}
    	Given Assumption \ref{assum2}, 
       suppose that there exists a constant $\kappa_2>0$ such that 
        $F(\boldsymbol{u})$ is  $\kappa_2$-strongly monotone.
        Then
		if  	
		$ H(\varLambda_{{TO}},\varLambda_{{NE}})<\eta$, 
        \begin{align}
            H(\Upsilon_{TO},\Upsilon_{NE})<\frac{\eta}{\kappa_2}. \nonumber
        \end{align} 
\end{mythm}

In Theorem \ref{c7}, the upper bound on the Hausdorff metric  $H(\Upsilon_{TO},\Upsilon_{NE})$ is mainly affected by the Hausdorff metric $H(\varLambda_{TO},\varLambda_{NE})$. Particularly, when the NE $\boldsymbol{u}^{\Diamond}$ is an interior point of the strategy set $\boldsymbol{\Xi}$, the set $\varLambda_{{NE}}$ is the zero set, then
the Hausdorff metric on the gradient space becomes  $ H(\varLambda_{{TO}},\varLambda_{{NE}})=\sup\nolimits_{\boldsymbol{u}^{*}\in \varLambda_{{TO}}} \|F(\boldsymbol{u}^{*})\| $. Regarding the Hausdorff metric $H(\varLambda_{TO},\varLambda_{NE})$ as a perturbation measure, a small value  yields a lower bound. Moreover, the Hausdorff metric characterizes the maximum deviation between two sets in the worst case. If the bound is sufficiently small, then even at the farthest point, the NE set remains close to the team-optimal solution set.

\vspace{-0.2cm}
\section{Coincidence of team-optimal solution and NE } 

In this section, we consider a special type of non-cooperative game, namely the potential game, and investigate the coincidence relationship between the team-optimal solution and the NE.


We define the potential game $\{\mathcal{N}, \{\Xi_i\}_{i\in\mathcal{N}}, \{\Tilde{\mathcal{C}}_i\}_{i\in\mathcal{N}}\}$. The first two tuples 
are the same as the game introduced in Section  \ref{Non-cooperative game}, but the payoff function
$\Tilde{\mathcal{C}}_i(u_i,\boldsymbol{u}_{-i})$ of each DM differs. Here,
$\Tilde{\mathcal{C}}_i(u_i,\boldsymbol{u}_{-i})$
is defined as each DM's marginal contribution to the overall  transmission cost, 
i.e.,
\begin{align}\label{eqs51}
&\Tilde{\mathcal{C}}_i(u_i,\boldsymbol{u}_{-i})=\sum_{l=1}^L\; \rho_{l} \omega(l,  i) \left[\sum_{k=1}^{P_i} \;
f_l(z_l(u_i,\boldsymbol{u}_{-i}))\right].
\end{align}

Then, we introduce the concept of potential game
\cite{monderer1996potential}. 
\begin{mydef}[Potential game]\label{d3}
	A game $\{\mathcal{N}, \!\{\Xi_i\}_{i\in\mathcal{N}}, $ $ \!\{\Tilde{\mathcal{C}}_i\!\}_{i\in\!\mathcal{N}}\}$ is a potential
	game if there exists a potential function $\Phi$ such that for  $\boldsymbol{u}\in \boldsymbol{\Xi}$, $i\in\mathcal{N}$,  and  $u_i^{\prime}\in \Xi_i$,
       \vspace{-0.15cm}
	\begin{equation}\label{pp1}
	\Phi(u_{i}^{\prime}, \boldsymbol{u}_{-i})-\Phi\left(\boldsymbol{u}\right)=\!\Tilde{\mathcal{C}}_i(u_{i}^{\prime}, \boldsymbol{u}_{-i})-\Tilde{\mathcal{C}}_i\left(\boldsymbol{u}\right). 
    \vspace{-0.15cm}
	\end{equation}
\end{mydef}

It follows from Definition   \ref{d3} that 
any unilateral deviation from a strategy profile results in the same change in both individual payoffs and a unified potential function \cite{monderer1996potential,xu2024multi}.
The following result verifies that $\mathcal{C}(\boldsymbol{u})$  is a potential function. The proof is presented in Appendix  \ref{lea5}.   
\begin{proposition}\label{p1} 
     Given the function $\!\mathcal{C}\!$  and payoffs $\Tilde{\mathcal{C}}_i$ for $i\!\in\!\mathcal{N}$,  the   game $\{\mathcal{N}, \{\Xi_i\}_{i\in\mathcal{N}}, \{\Tilde{\mathcal{C}}_i\}_{i\in\mathcal{N}}\}$ is a  potential  game.
\end{proposition}

The following result provides a consistent relationship between the team-optimal solution and the NE.
The proof is presented in Appendix \ref{lea9}.
\begin{proposition}\label{t89}
Under potential game, if $\boldsymbol{u}^*$ is a team-optimal solution minimizing $\mathcal{C}$, then $\boldsymbol{u}^*$ is an  NE $\boldsymbol{u}^\Diamond$ of the game $\{\mathcal{N}, \!\{\Xi_i\}_{i\in\mathcal{N}}, \!\{\Tilde{\mathcal{C}}_i\}_{i\in\mathcal{N}}\}$. Moreover, if Assumption \ref{assum2} holds, then $\boldsymbol{u}^\Diamond$ is also a team-optimal solution $\boldsymbol{u}^*$.
\end{proposition}

Proposition~\ref{t89} establishes the consistency between the team-optimal solution and the NE. 
When the condition in Proposition~\ref{t89} is satisfied, the Hausdorff metric between the team-optimal solution set and the NE set is zero. 
This indicates that the cost for the whole team remains unchanged. Even if some DMs choose the NE strategy,  the remaining DMs can still follow the team-optimal strategies.

\begin{table*}[t]
	\centering
	\footnotesize
     \vspace{-0.15cm}
	\caption{Five different traffic scenarios.}
		\setlength\tabcolsep{12pt}
		\renewcommand\arraystretch{1.05}
		\begin{tabular}{c|c|c|c|c|c|c|c|c|c|c|c}
			\hline
			\hline
			\specialrule{0em}{0.2pt}{0.5pt}
  &$ \rho_1 $ &$\rho_2 $  &$ \rho_3 $& {$ \rho_4 $}&  $ \rho_5 $ &$ \rho_6 $ &$\rho_7 $  &$ \rho_8 $& $ \rho_9 $&  $ \rho_{10} $ &  $ \rho_{11} $ \\ 
   \hline	
            \rowcolor{mygray}
             {Case 1}& 10 & 11  &15 & 8 & 6 & 7 & 5  &13 & 12 &  11  &  9\\
			\hline
            {Case 2}& 12 & 9  &14 & 7 &  6 & 5 & 6  &14 & 10 &  8  &  12\\
			\hline
            {Case 3}& 18 & 12  &17 & 5 &  7 & 6 & 8  &10 & 8 &  9  &  10\\
			\hline
           {Case 4}& 14 & 10  &16 & 6 &  5 & 8 & 7  &12 & 9 &  10  &  11\\
			\hline
            {Case 5}& 16 & 13  &12 & 6 & 5 & 6 & 7  &15 & 11 &  8  &  13\\
			\hline\hline
	\end{tabular}
     \vspace{-0.35cm}
	\label{tab6435}
\end{table*}

\section{Simulations Studies}\label{s3} 

In this section, we illustrate the relationship between
the team-optimal solution and the NE in two logistics transportation examples.

\begin{figure}[ht]
	\centering	
     \vspace{-0.2cm}
	\includegraphics[scale=0.11]{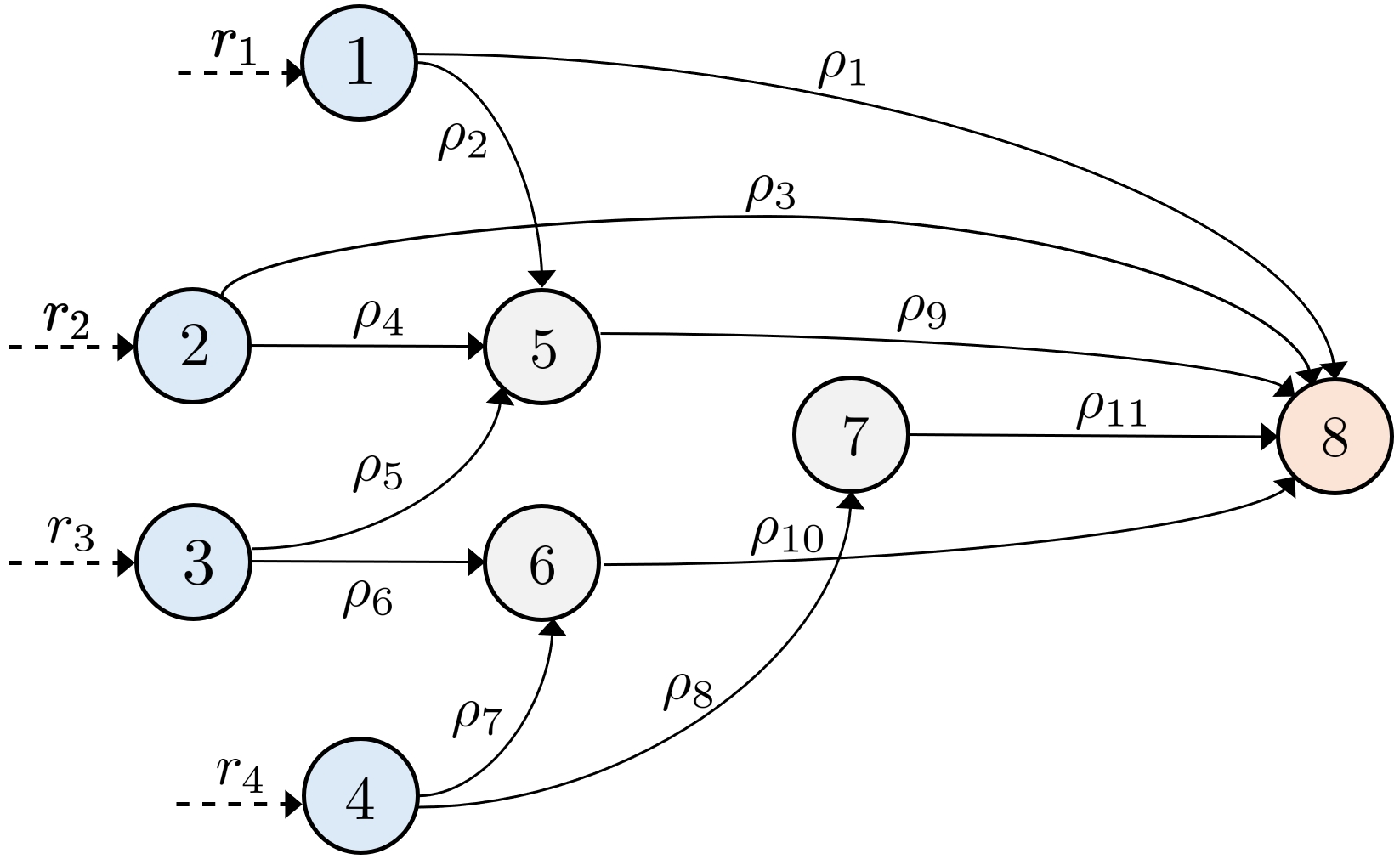}\\
 \vspace{-0.2cm}
	\caption{Logistics network with four DMs.
    }
	\label{fig17}
	\vspace{-0.3cm}
\end{figure}

Consider a logistics transportation example with four shippers acting as DMs.  As shown in Fig.~\ref{fig17}, each shipper has two available paths.  Shipper 1 can choose between paths 1-8 and 1-7-8.
Shipper 2 can choose between paths 2-8 and 2-5-8.
Shipper 3 can choose between paths 3-5-8 and 3-6-8.
Shipper~4 can choose between paths 4-6-8 and 4-7-8.
Each shipper needs to transport a specific amount of goods. According to the general load capacity limits, take $r_1=10$ tons, 
$r_2=15$ tons, $r_3=8$ tons, and $r_4=12$ tons. The transportation network consists of 11 edges with congestion cost coefficients $\rho_l$, representing the transportation cost per unit
flow, measured in $\$/\operatorname{ton}$. 
The specific values of $\rho_l$ are listed in Case 1 of Table~\ref{tab6435}, where the short-distance transportation costs ($\rho_4$, $\rho_5$, $\rho_6$, $\rho_7$) are generally lower than the long-distance transportation costs ($\rho_1$,  $\rho_2$, $\rho_3$,  $\rho_8$,  $\rho_9$, $\rho_{10}$,  $\rho_{11}$).
Besides, 
we set the transmission cost $f_l(z_l)$ on edge $l$ as quadratic, and the cost per unit traffic $g_l(z_l)$  as linear, where $a_l=0$ and $b_l=1$.

        \begin{figure}[h]
	\centering	
	\begin{subfigure}
		\centering
		\includegraphics[width=4.4cm]{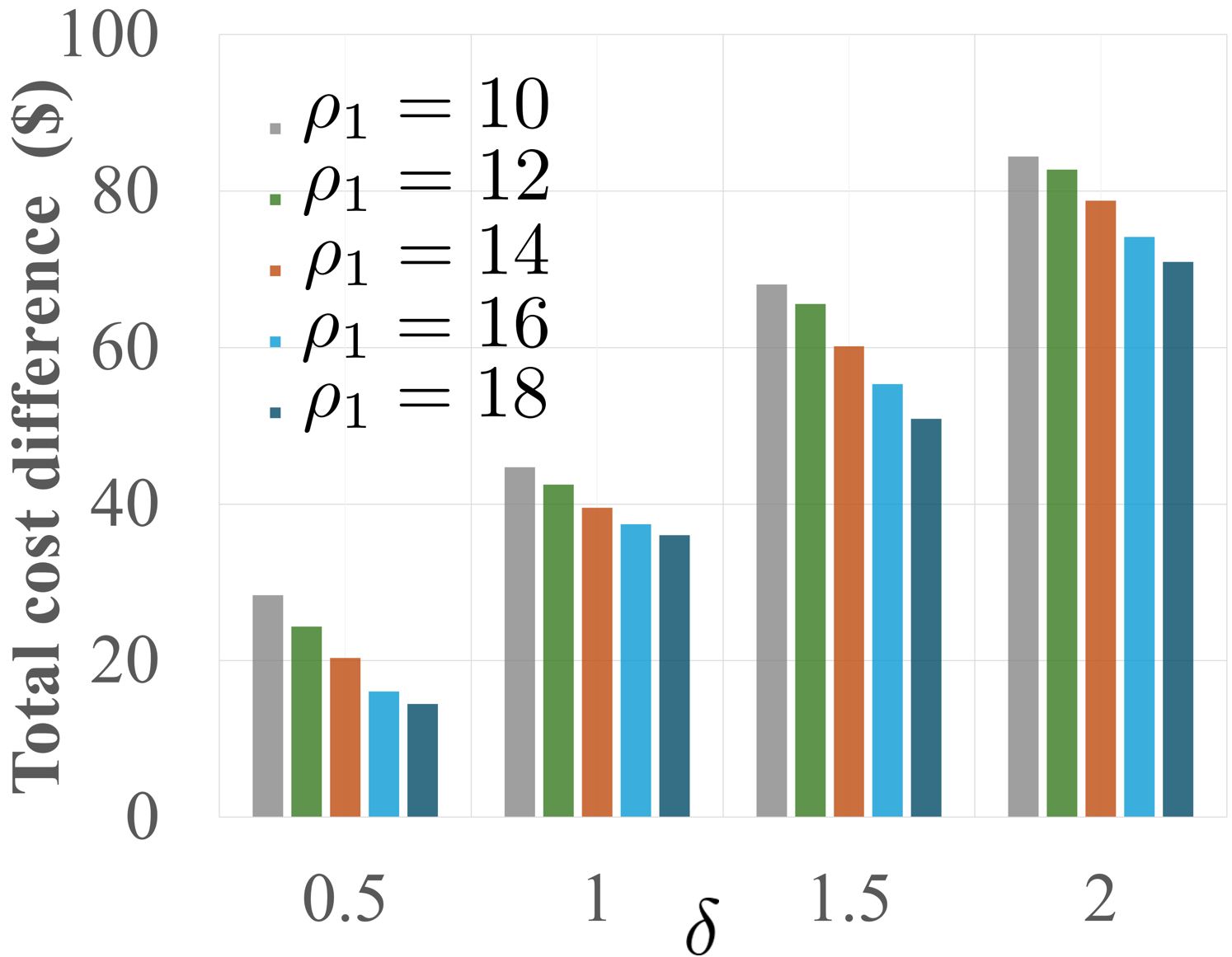}
	\end{subfigure}%
	%
	\begin{subfigure}
		\centering
		\includegraphics[width=4.3cm]{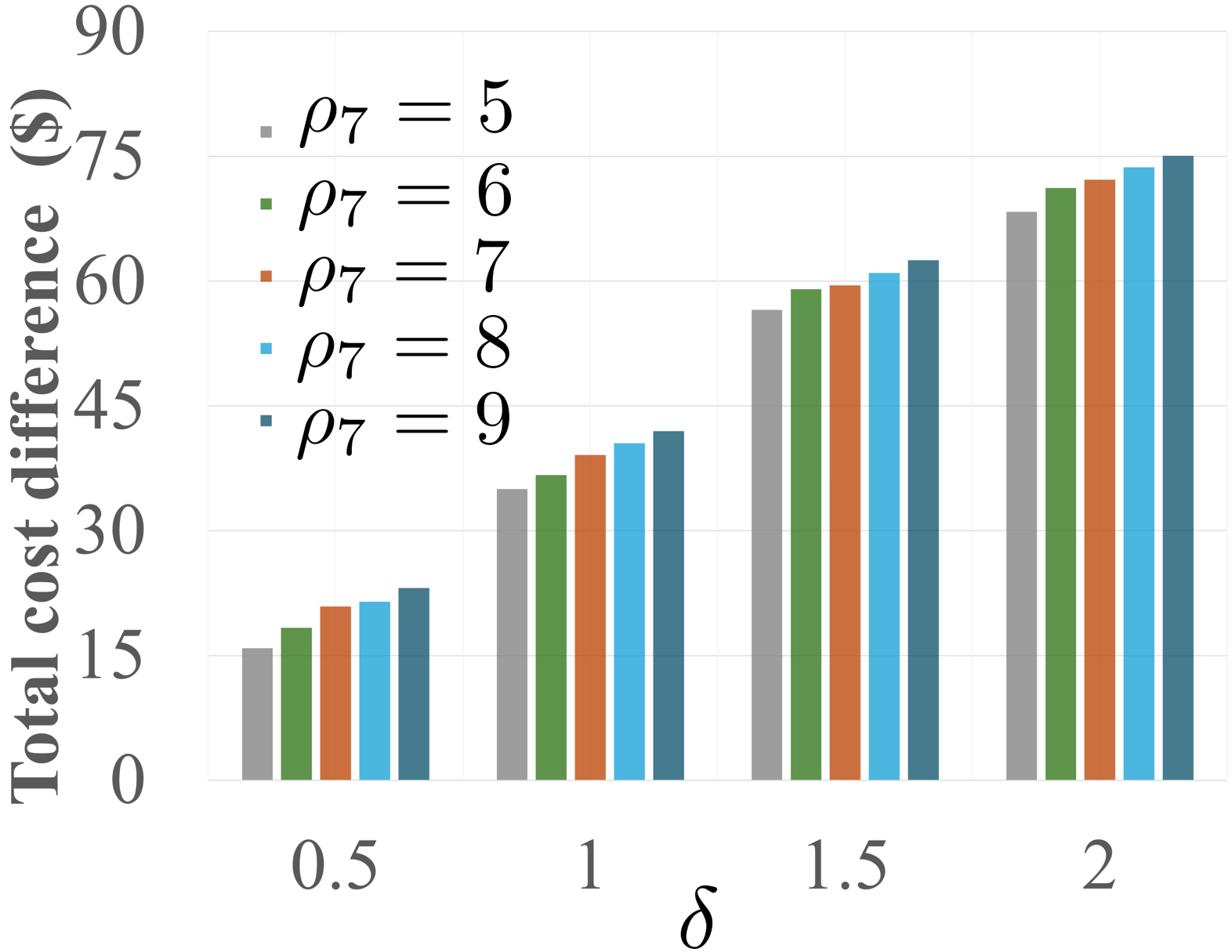}
	\end{subfigure}%
	\centering
	\vspace{-0.6cm}
	\caption{Deviation between the team-optimal solution and NE.} 
	\label{fig44}
	\vspace{-0.2cm}
\end{figure}

We first focus on the deviation bound between the unique team-optimal solution and the unique NE. From \eqref{totalcost},
the strong monotonicity constant $\kappa_1$ is mainly influenced by \( \rho_l \) for $l= 1,\dots, 11$. 
This further affects the deviation bound \(\frac{ \delta}{\kappa_1 s} \) 
in Theorem~\ref{t7}. 
Based on Case 1, we randomly select two weight coefficients, \( \rho_1 \) and \( \rho_7 \), representing the long-distance path cost and the short-distance path cost, respectively. We then observe how changes in these parameters influence the deviation bound
in Fig.~\ref{fig44}.
The horizontal axis represents the value of $\delta$ in
Theorem~1. The vertical axis represents the difference between the total transmission cost  $\mathcal{C}(\boldsymbol{u}^*)$ and $\mathcal{C}(\boldsymbol{u}^{\Diamond})$, corresponding to the 
 value of \(\frac{ \delta}{\kappa_1 s} \).
Fig.~\ref{fig44} shows that increasing the long-distance cost \( \rho_1 \) narrows the total-cost gap, whereas increasing the short-distance cost \( \rho_7 \) enlarges it.
A smaller value of $\delta$ further reduces the gap, indicating that when the NE approaches the team optimum, network efficiency changes little.


\begin{figure}[ht]
	\centering	
     \vspace{-0.2cm}
	\includegraphics[scale=0.14]{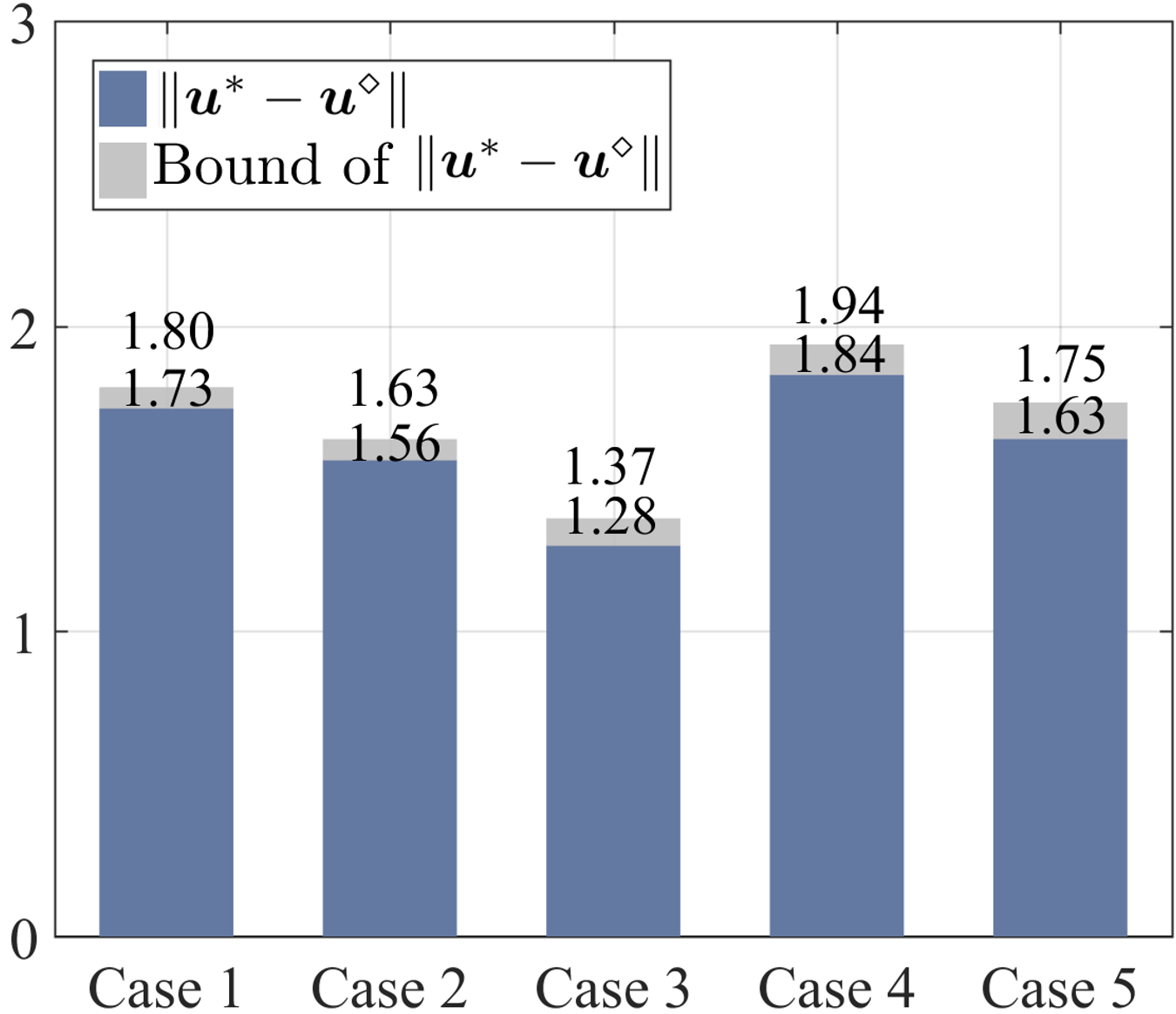}\\
 \vspace{-0.2cm}
	\caption{Deviation bound between team-optimal solution and NE.}
	\label{fig32}
	\vspace{-0.3cm}
\end{figure}

Furthermore, in Table \ref{tab6435}, we randomly generate  four additional scenarios with different values  of $\rho_1,\dots,\rho_{11}$.   The short-distance transportation costs 
are set between 5 and 10 $\$/\operatorname{ton^2}$, while the long-distance transportation costs 
are set between 8 and 20 $\$/\operatorname{ton^2}$. Fig.~\ref{fig32}  presents the   bound  \(\frac{ \delta}{\kappa_1 s} \) 
for these cases. In Fig.~\ref{fig32}, the deviation $\|\boldsymbol{u}^*-\boldsymbol{u}^{\Diamond}\|$ remains consistently within the corresponding upper bound, and the difference between them is relatively small.

\begin{figure}[ht]
	\centering	
     \vspace{-0.2cm}
	\includegraphics[scale=0.1]{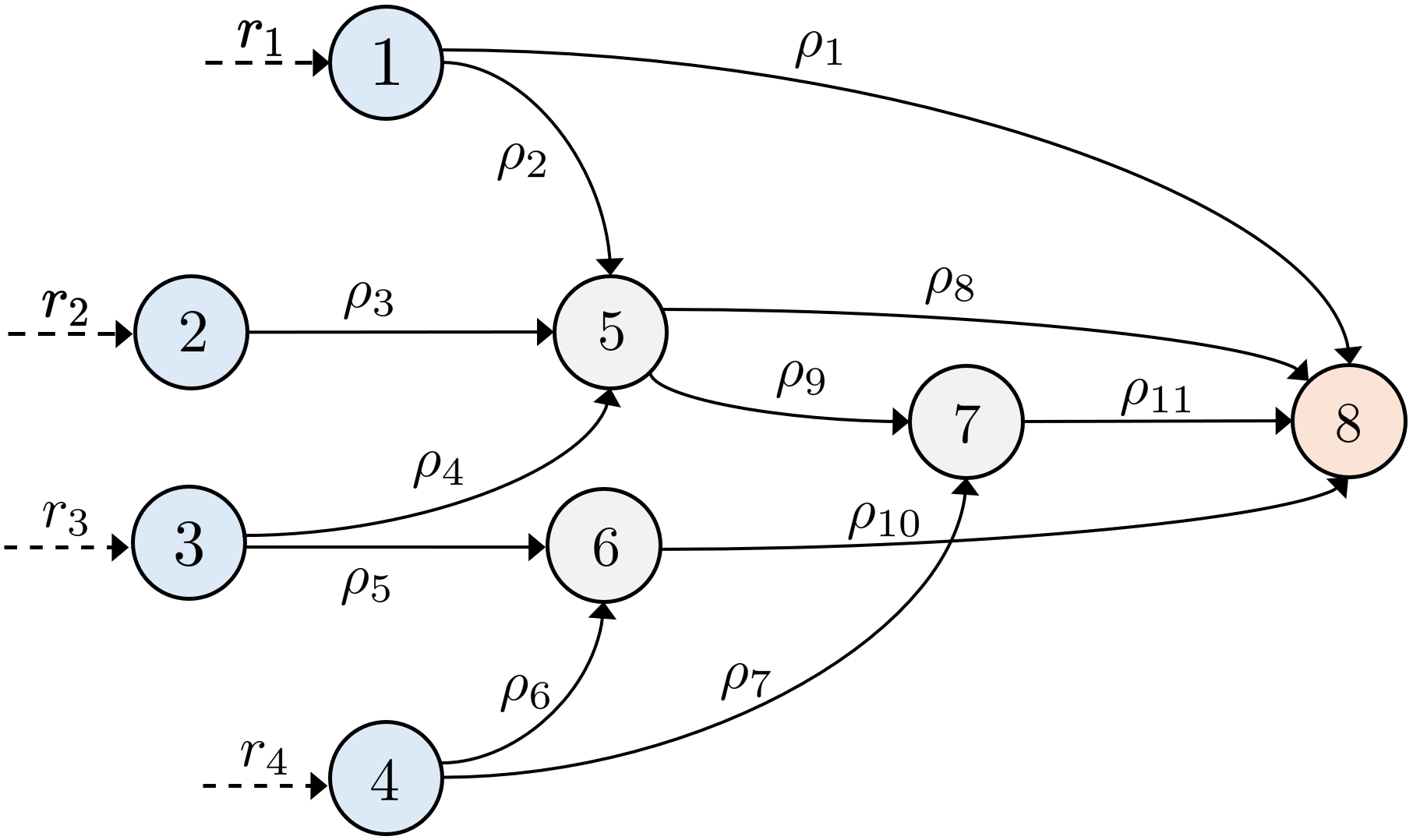}\\
 \vspace{-0.2cm}
	\caption{Another logistics network with four DMs.
    }
	\label{fig12}
	\vspace{-0.3cm}
\end{figure}

\begin{figure}[t]
   \vspace{-0.2cm}
			\hspace{-0.3cm}
			\centering	
			\subfigure[Non-cooperative game]{
				\begin{minipage}[t]{0.49\linewidth}
					\centering
					\includegraphics[width=4.6cm]{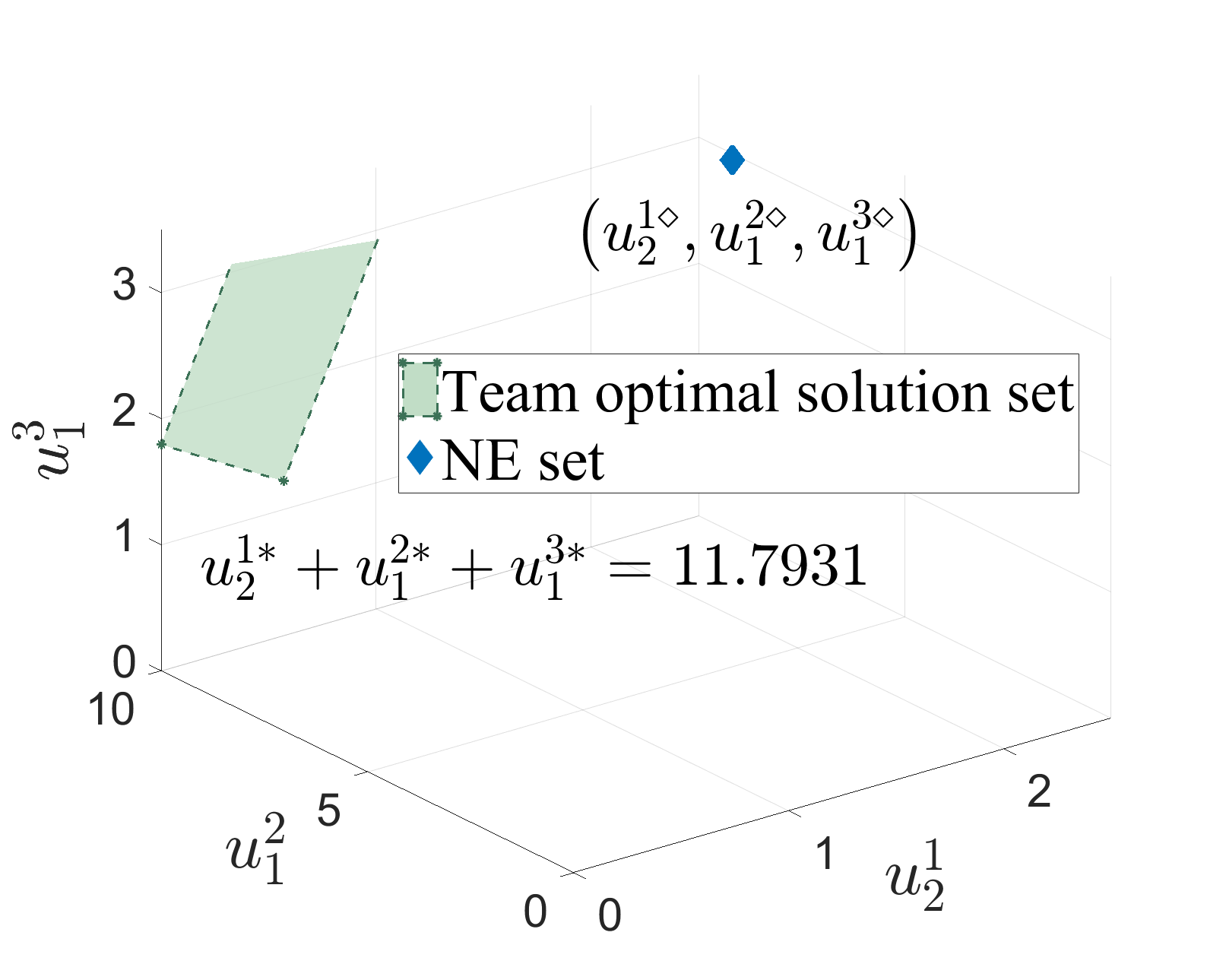}
				\end{minipage}%
			}%
			\hspace{-0.2cm}
			\subfigure[Potential game]{
				\begin{minipage}[t]{0.49\linewidth}
					\centering
					\includegraphics[width=4.6cm]{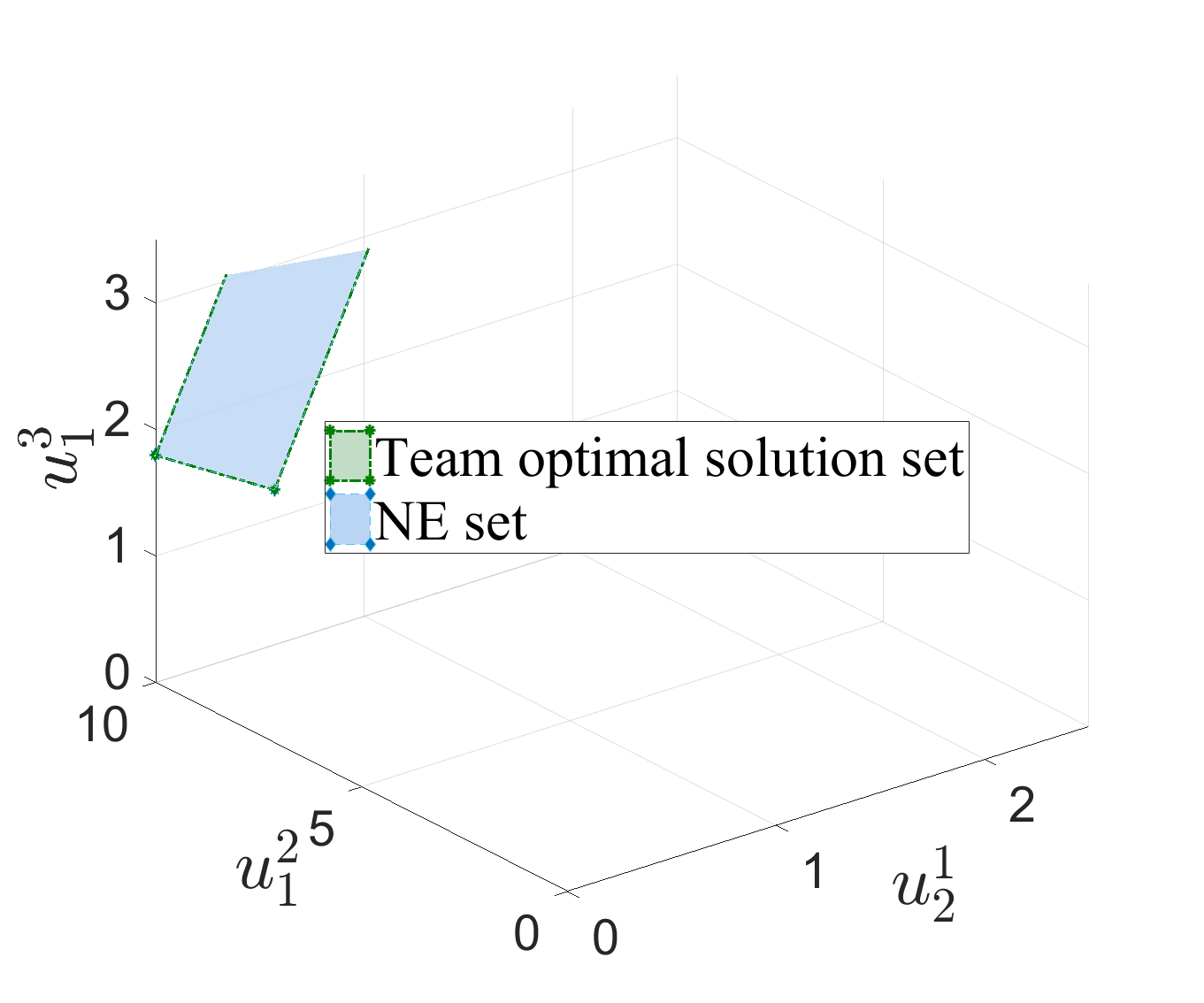}
				\end{minipage}%
			}%
			\centering
            \vspace{-0.25cm}
			\caption{Team-optimal solution and NE in different settings. Any team-optimal solution is an NE in the potential game.} 
   \vspace{-0.2cm}
			\label{fig54}
		\end{figure}

Next, let us consider another transportation network, shown in Fig.~\ref{fig12}. Based on this network, we study the relationship between the team-optimal solution set and the NE set in both the non-cooperative game and the potential game.  In this example, DM$_1$ has three available paths, DM$_2$ has two, DM$_3$ has three, and DM$_4$ has two. Let $\rho_1=10$,  $\rho_2=8$, $\rho_3=6$, $\rho_4=7$, $\rho_5=8$, $\rho_6=10$, $\rho_7=5$, $\rho_8=8$,  $\rho_9=7$, $\rho_{10}=6$,  $\rho_{11}=10$. 
We compute the team-optimal solution via  \eqref{team_function},   the NE  in the non-cooperative game via  \eqref{payoff1}, and the NE in the potential game via \eqref{eqs51}. To visualize the strategy space of all DMs, we map it to the strategy space of $u_2^1$, $u_1^2$, $u_1^3$. As shown in Fig.~\ref{fig54}, the strategy profile $\boldsymbol{u}^*$ that satisfies the  constraints $\boldsymbol{\Xi}$ and the  condition $u_2^{1*}+u_1^{2*}+u_1^{3*}=11.7931$ is a team-optimal solution. In Fig.~\ref{fig54}(a), a team-optimal solution is not an NE  in the non-cooperative game setting. In Fig. \ref{fig54}(b),  any team-optimal solution is an NE in the potential game setting.

\begin{figure}[ht]
\vspace{-0.2cm}
	\centering	
	\includegraphics[scale=0.14]{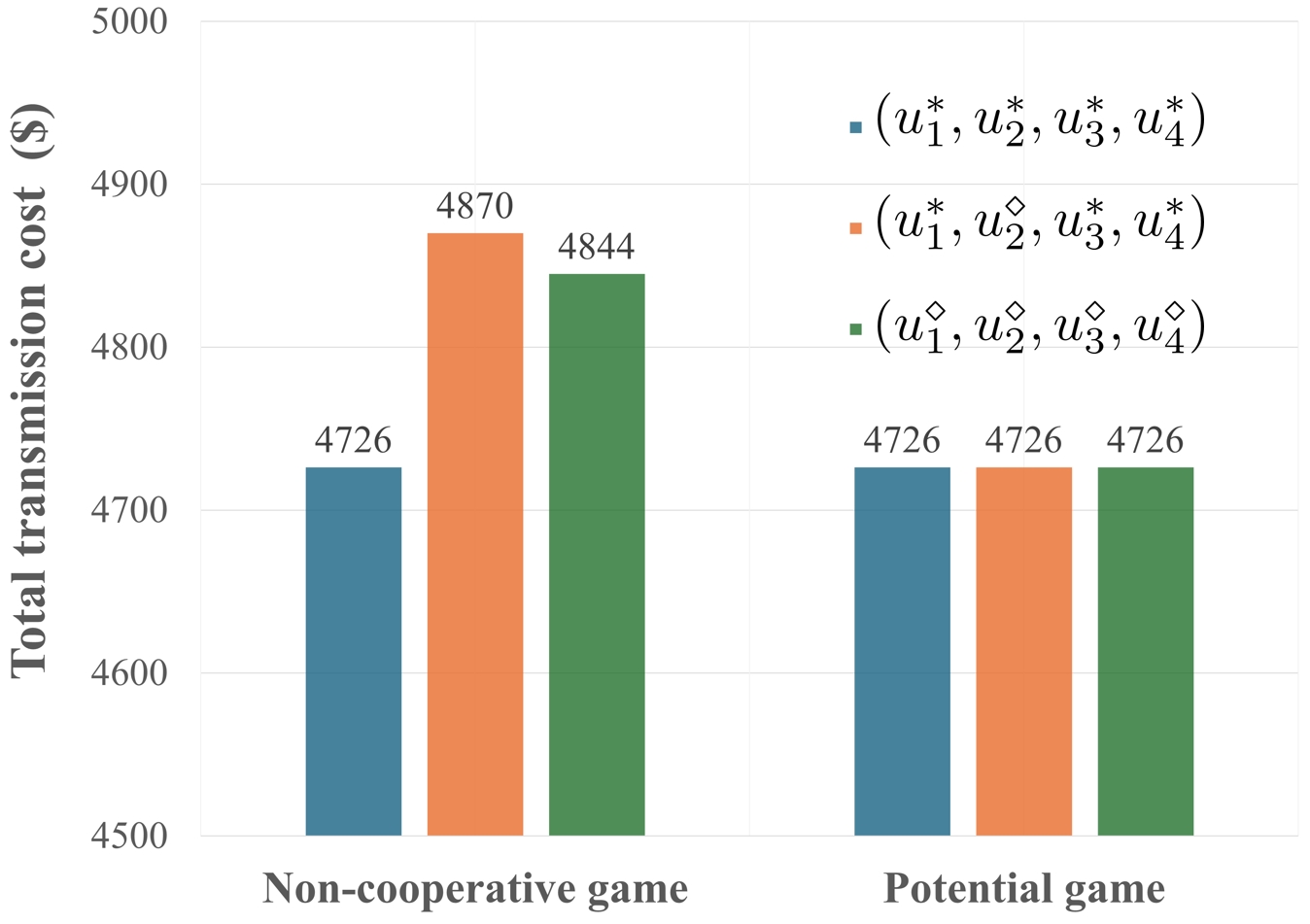}\\
 \vspace{-0.2cm}
	\caption{Total cost $\mathcal{C}$ corresponding to different strategy profiles.}
	\label{fig22}
\end{figure}

In this example, the individual cost for \( \text{DM}_2 \) under the non-cooperative game setting with the NE strategy is lower than that under the team-optimal solution, which motivates \( \text{DM}_2 \) to deviate. Moreover, this deviation has the greatest impact on the difference between the team-optimal solution set and the NE set.
Therefore, we consider  three strategy  profiles: (i) all DMs choose the team-optimal strategy;
(ii) DM$_2$ takes the NE strategy, while the others adopt the team-optimal strategies; (iii) all DMs choose the NE strategies.  Fig.~\ref{fig22} investigates how different strategy profiles affect the total cost $\mathcal{C}$.
In a general non-cooperative game, the misalignment between individual objectives and the collective goal leads to a discrepancy between the NE and the team-optimal solution. The ideal scenario occurs when all DMs adopt the team-optimal strategies. 
However, when DM$_2$ deviates from the NE strategy, the resulting strategy profile does not form an equilibrium, increasing the total cost. 
When all DMs adopt the NE strategy, the cost decreases relatively. This suggests that the remaining DMs need to make a trade-off between the team-optimal strategy and the NE strategy.  
In the potential game setting, individual interests align with the collective objective, making the NE identical to the team-optimal solution. Therefore, even if DM$_2$ deviates,
the total cost remains unchanged, allowing the remaining DMs to still adopt the team-optimal strategies.

\section{Concluding Remarks}\label{s3} 
In this paper, we addressed a flow assignment problem and investigated the relationship between the team-optimal solutions and the NE to evaluate the impact of unilateral deviations on team performance. {We studied the upper bound on the deviation between the team-optimal solution and the NE and established the consistency between the team-optimal solution and the NE in a potential game.}
Finally, we presented the applicability of our results in logistics examples. 

A potential direction of future research could consider investigating these questions in the case of random flow assignments. 



\appendix

\subsection{Proof of   Theorem~\ref{t7}}\label{lea8} 
Before delving into the details, we present our proof idea. We establish a deviation bound between the team-optimal solution and the NE using gradient information. To accomplish this, we propose two projected gradient algorithms for finding the team-optimal solution and the NE. Subsequently, we treat the gradient difference as a perturbation term and derive the deviation bound based on the stability of perturbed systems.

Let us consider the following  algorithm for seeking the team-optimal solution, i.e, for $i\in\mathcal{N}$, 
\begin{equation}\label{team1}
\begin{aligned}
    &\dot{y}_i=-\nabla_{{u}_i}\mathcal{C}(\boldsymbol{u})+{u}_i-y_i\\
    &u_i=\Pi_{{\Xi}_i}(y_i)
    \end{aligned}
\end{equation}
Here,  $\Pi_{{\Xi}_i}:\mathbb{R}^{P_i}\rightarrow {\Xi}_i$ is the projection map,  defined as $\Pi_{{\Xi}_i}(x)=\operatorname{argmin}_{y\in {\Xi}_i} \|x-y\|$.
The compact form of \eqref{team1} can be written as
\begin{equation}\label{dynamics_team}
\begin{aligned}
 &\dot{\boldsymbol{y}}=-G(\boldsymbol{u})+\boldsymbol{u}-\boldsymbol{y}\\
    &\boldsymbol{u}=\Pi_{\boldsymbol{\Xi}}(\boldsymbol{y})
  \end{aligned}
\end{equation}
It is clear that the equilibrium of dynamics \eqref{dynamics_team} is $\boldsymbol{u}^*$.  Then we show that \eqref{dynamics_team} converges at an exponential rate.
Consider the  candidate Lyapunov function $V(\boldsymbol{u})=\frac{1}{2}\|\boldsymbol{u}-\boldsymbol{u}^{*}\|^2$. 
Its time derivative along the trajectory of \eqref{dynamics_team} is
\begin{equation*}
    \begin{aligned}
        \dot{V}(\boldsymbol{u})&=(\boldsymbol{u}-\boldsymbol{u}^*)^T\dot{\boldsymbol{y}}\\
    &=(\boldsymbol{u}\!-\!\boldsymbol{u}^*)^T(-G(\boldsymbol{u})+\boldsymbol{u}-\boldsymbol{y})\!\\
    &=-(\boldsymbol{u}\!-\!\boldsymbol{u}^*)^TG(\boldsymbol{u})+(\boldsymbol{u}\!-\!\boldsymbol{u}^*)^T(\boldsymbol{u}-\boldsymbol{y})\\
    &\leq -(\boldsymbol{u}\!-\!\boldsymbol{u}^*)^TG(\boldsymbol{u})\\
    &=-(\boldsymbol{u}\!-\!\boldsymbol{u}^*)^T(G(\boldsymbol{u})-G(\boldsymbol{u}^*))-(\boldsymbol{u}\!-\!\boldsymbol{u}^*)^TG(\boldsymbol{u}^*)\\
    &\leq -(\boldsymbol{u}\!-\!\boldsymbol{u}^*)^T(G(\boldsymbol{u})-G(\boldsymbol{u}^*))\\
    &\leq-\kappa_1 \|\boldsymbol{u}-\boldsymbol{u}^*\|^2
\end{aligned}
\end{equation*}
Then we have
\begin{align}
     \dot{V}(\boldsymbol{u})\leq -\kappa_1 \|\boldsymbol{u}-\boldsymbol{u}^*\|^2\leq -2\kappa_1 V(\boldsymbol{u}).\nonumber
\end{align}
This represents the exponential convergence rate.

Similarly, 
consider the compact form of the NE-seeking dynamics
\begin{equation}\label{dynamics_ne}
\begin{aligned}
 &\dot{\boldsymbol{y}}=-F(\boldsymbol{u})+\boldsymbol{u}-\boldsymbol{y}\\
    &\boldsymbol{u}=\Pi_{\boldsymbol{\Xi}}(\boldsymbol{y})
  \end{aligned}
\end{equation}
We can similarly prove that \eqref{dynamics_ne} converge to $\boldsymbol{u}^{\Diamond}$
with an exponential convergence rate. 
Then, 
let us  rewrite the NE-seeking dynamics \eqref{dynamics_ne} as 
\begin{align}\label{perturb}
   &\dot{\boldsymbol{y}}=-G(\boldsymbol{u})+\boldsymbol{u}-\boldsymbol{y}+(G(\boldsymbol{u})-F(\boldsymbol{u}))\\
    &\boldsymbol{u}=\Pi_{\boldsymbol{\Xi}}(\boldsymbol{y})
\end{align}
which can be regarded a perturbed system of \eqref{dynamics_team}. 
Moreover, since $\boldsymbol{\Xi}$ is compact, there exists a constant $\delta$ such that $\| F(\boldsymbol{u})- G(\boldsymbol{u})\|\leq\delta$.
Then the Lyapunov function $V(\boldsymbol{u})$ along the trajectories
of \eqref{perturb} satisfies 
\begin{equation*}
    \begin{aligned}
    \dot{V}(\boldsymbol{u})
    &\leq -\kappa_1\|\boldsymbol{u}-\boldsymbol{u}^*\|^2+\|\frac{\partial{V}}{\partial \boldsymbol{u}}\|\| F(\boldsymbol{u})- G(\boldsymbol{u})\|\\
     &\leq -\kappa_1\|\boldsymbol{u}-\boldsymbol{u}^*\|^2+ \delta \|\boldsymbol{u}-\boldsymbol{u}^*\|\\
     &=-(1-s)\kappa_1\|\boldsymbol{u}-\boldsymbol{u}^*\|^2-s \kappa_1\|\boldsymbol{u}-\boldsymbol{u}^*\|^2+\delta \|\boldsymbol{u}-\boldsymbol{u}^*\|\\
     &\leq -(1-s)\kappa_1\|\boldsymbol{u}-\boldsymbol{u}^*\|^2, \quad \forall \|\boldsymbol{u}-\boldsymbol{u}^*\|\geq \delta/s\kappa_1.
\end{aligned}
\end{equation*}
Referring to \cite[Theorem 4.18]{khalil2002nonlinear}, 
when $V(\boldsymbol{u})\geq\frac{1}{2}(\delta/s\kappa_1)^2$, we have
$\|\boldsymbol{u}-\boldsymbol{u}^*\|\geq \delta/s\kappa_1$ and $\dot{V}(\boldsymbol{u})\leq  -2(1-s)\kappa_1 V(\boldsymbol{u})$. This implies
\begin{align*}
\|\boldsymbol{u}-\boldsymbol{u}^*\|^2 &\leq \sqrt{2V(\boldsymbol{u}(t_0))}e^{-(1-s)\kappa_1(t-t_0)} \\
&\leq e^{-(1-s)\kappa_1(t-t_0)}\|\boldsymbol{u}(t_0)-\boldsymbol{u}^*\|
\end{align*}
over the interval $[t_0,t_0+T)$ for some finite $T$. For $t\geq t_0+T$, we have
\begin{align*}
    \|\boldsymbol{u}-\boldsymbol{u}^*\|\leq \sqrt{2V(\boldsymbol{u})} \leq \delta/s\kappa_1,
\end{align*}
and the proof is complete.
\hfill $\square$

\subsection{Proof of   Theorem~\ref{c7}}\label{lea3} 
It follows from $F$ being $\kappa_2$-strongly monotone in $\boldsymbol{u}$, we have 
\begin{align*}
    \kappa_2 \|\boldsymbol{u}^{*}-\boldsymbol{u}^{\Diamond} \|\leq \|F(\boldsymbol{u}^{*})-F(\boldsymbol{u}^{\Diamond}) \|.
\end{align*}
Then, following  the definition of  $H(\varLambda_{{TO}},\varLambda_{{NE}})$, we get 
\begin{align*}
	&\quad\quad H(\varLambda_{{TO}},\varLambda_{{NE}})\\
	&= \max\Big\{\!\!\sup_{F(\boldsymbol{u}^{*})\in \varLambda_{{TO}}}\!\inf_{F(\boldsymbol{u}^{\Diamond})\in \varLambda_{{NE}}}\!\|\!F(\boldsymbol{u}^{*})\!-\!F(\boldsymbol{u}^{\Diamond})\!\|,\\
	&\quad\quad\quad\quad\sup\limits_{F(\boldsymbol{u}^{\Diamond})\in \varLambda_{{NE}}} \!\inf_{F(\boldsymbol{u}^{*})\in \varLambda_{{TO}}} \!\!\|\!F(\boldsymbol{u}^{*})\!-\!F(\boldsymbol{u}^{\Diamond})\!\| \Big\}\\
	&\geq  \max\Big\{\!\!\sup_{F(\boldsymbol{u}^{*})\in \varLambda_{{TO}}}\inf_{F(\boldsymbol{u}^{\Diamond})\in \varLambda_{{NE}}}\kappa_2\!\|\boldsymbol{u}^{*}\!-\!\boldsymbol{u}^{\Diamond}\|\!,\\
	&\quad\quad\quad\quad\sup\limits_{F(\boldsymbol{u}^{\Diamond})\in \varLambda_{{NE}}} \inf \limits_{F(\boldsymbol{u}^{*})\in \varLambda_{{TO}}} \kappa_2\|\boldsymbol{u}^{*}\!-\!\boldsymbol{u}^{\Diamond}\!\| \Big\}
\end{align*}
Therefore,
\begin{align*}
		&\quad	H(\Upsilon_{TO},\Upsilon_{NE})\\
		&= \max\Big\{\sup_{\boldsymbol{u}^{*} \in \Upsilon_{TO} }\!\inf_{\boldsymbol{u}^{\Diamond} \in \Upsilon_{NE}}\|\boldsymbol{u}^{*}\!-\!\boldsymbol{u}^{\Diamond}\|,\\
        &\quad\quad\quad\quad\sup\limits_{\boldsymbol{u}^{\Diamond}\in \Upsilon_{NE}} \inf \limits_{\boldsymbol{u}^{*}  \in \Upsilon_{TO}} \|\boldsymbol{u}^{*}\!-\!\boldsymbol{u}^{\Diamond}\| \Big\}\\	
			&\leq   \frac{H(\varLambda_{{TO}},\varLambda_{{NE}})}{\kappa_2}\\
			&\leq\!\frac{\eta}{\kappa_2}. 
			\end{align*}
This completes the proof. \hfill $\square$

\subsection{Proof of   Proposition \ref{p1}}\label{lea5}  
Given that the payoff functions $\Tilde{\mathcal{C}}_{i}$ are continuously differentiable, it follows from \cite{monderer1996potential} that the condition in Definition~\ref{d2} is equivalent to 
\begin{align*}
    {\nabla_{u_i}\Tilde{\mathcal{C}}_{i}(u_{i}, \boldsymbol{u}_{-i})}={\nabla_{u_i} \mathcal{C}(u_{i}, \boldsymbol{u}_{-i})}
\quad \text { for } i\in\mathcal{N}.
\end{align*}
By computing the partial derivatives of $\Tilde{\mathcal{C}}_{i}$ and $\mathcal{C}$, respectively, we can verify that $\{\mathcal{N}, \{\Xi_i\}_{i\in\mathcal{N}}, \Tilde{\mathcal{C}}_i\}_{i\in\mathcal{N}}\}$ is a potential game. \hfill $\square$

\subsection{Proof of   Proposition \ref{t89}}\label{lea9} 

Recall the definition of team-optimal solution, $\boldsymbol{u}^*$ is said to be team-optimal if
\begin{align*}
    \mathcal{C}\left(u_{i}^*, \boldsymbol{u}_{-i}^{*}\right)\leq \mathcal{C}\left(u_{i}, \boldsymbol{u}_{-i}\right), \forall \boldsymbol{u}\in \boldsymbol{\Xi}.
\end{align*}
Then, for each $i\in\mathcal{N}$,  let $\boldsymbol{u}$ be the tuple whose $j$-th subvector
is equal to $u_{j}^*$ for $j \neq i$ and $i$-th subvector is equal to $u_i$, where $u_i$ is an
arbitrary element of the set $\Xi_i$.  We have
\begin{align*}
\mathcal{C}\left(u_{i}^*, \boldsymbol{u}_{-i}^{*}\right)\leq \mathcal{C}\left(u_{i}, \boldsymbol{u}_{-i}^{*}\right),
\forall i \in \mathcal{N}, \forall u_{i}\in \Xi_{i}.
\end{align*}
This implies $\boldsymbol{u}^{*}$ is an NE that satisfies
\begin{align*}
\Tilde{\mathcal{C}}_i\left(u_{i}^{*}, \boldsymbol{u}_{-i}^{*}\right) \leq \Tilde{\mathcal{C}}_i\left(u_{i}, \boldsymbol{u}_{-i}^{*}\right),\;\forall i \in \mathcal{N}, \forall u_{i}\in \Xi_{i}.
\end{align*}

On the other hand, denote $\Tilde{F}(\boldsymbol{u})=\operatorname{col}\{\nabla_{{u_1}}\Tilde{\mathcal{C}}_1(\cdot,u_{-1}),\dots, \nabla_{{u_N}}\Tilde{\mathcal{C}}_N(\cdot,u_{-N})\}$ as the pseudo-gradient of potential game  $\{\mathcal{N}, \!\{\Xi_i\}_{i\in\mathcal{N}}, $ $ \!\{\Tilde{\mathcal{C}}_i\}_{i\in\mathcal{N}}\}$.  Since ${\nabla_{u_i}\Tilde{\mathcal{C}}_{i}(u_{i}, \boldsymbol{u}_{-i})}={\nabla_{u_i} \mathcal{C}(u_{i}, \boldsymbol{u}_{-i})}$, we have $\Tilde{F}(\boldsymbol{u})=G(\boldsymbol{u})$.  
Then, based on Assumption \ref{assum2},
the existence of NE $\boldsymbol{u}^{\Diamond}$ is guaranteed. Moreover, since $\Xi_i$ is compact and convex, $\Tilde{\mathcal{C}}_i(u_i, \boldsymbol{u}_{-i})$ is convex and continuously differentiable in $u_i$, by convexity and the minimum principle, $\boldsymbol{u}^{\Diamond}$  satisfies
\begin{align*}
(u_i-u_{i}^{\Diamond})^T\nabla_{u_i} \Tilde{\mathcal{C}}_i(\boldsymbol{u}^{\Diamond})\geq 0,\forall i \in \mathcal{N},  \forall u_i\in \Xi_i.
\end{align*}
Then by concatenating these inequalities,
\begin{align*}
(\boldsymbol{u}-\boldsymbol{u}^{\Diamond})^T\Tilde{F}(\boldsymbol{u}^{\Diamond})\geq 0, \forall \boldsymbol{u}\in \boldsymbol{\Xi},
\end{align*}
which implies
\begin{align*}
(\boldsymbol{u}-\boldsymbol{u}^{\Diamond})^TG(\boldsymbol{u}^{\Diamond})\geq 0, \forall \boldsymbol{u}\in \boldsymbol{\Xi},
\end{align*}
Thus, $\boldsymbol{u}^{\Diamond}$ is also a team-optimal solution that satisfies 
\begin{align*}
\mathcal{C}(\boldsymbol{u}^{\Diamond})\leq  \mathcal{C}(\boldsymbol{u}), \forall \boldsymbol{u}\in \boldsymbol{\Xi}.
\end{align*}
\hfill $\square$

\bibliographystyle{IEEEtran}
\bibliography{reference,IDSLab}

\end{document}